\documentclass[aps,pre,reprint,groupedaddress,showpacs,showkeys]{revtex4-1}

\usepackage{latexsym,amssymb,amsfonts,mathrsfs,xcolor,graphicx,amsmath}
\usepackage[american]{babel}
\usepackage[normalem]{ulem}

\newcommand{\az}[1]{\textcolor{black}{#1}}
\newcommand{\azz}[1]{\textcolor{black}{#1}}
\newcommand{\al}[1]{\textcolor{black}{#1}}

\begin{document}


\title{Asymmetric bistability of chiral particle orientation in viscous shear flows}


\author{Andreas Z\"ottl$^{a,b,c}$}
\thanks{AZ and FT contributed equally to this work.}
\author{Francesca Tesser$^{a}$} 
\thanks{AZ and FT contributed equally to this work.}
\author{Daiki Matsunaga$^d$}
\author{Justine Laurent$^a$}
\author{Olivia Du Roure$^{a}$}

\date{\today}

\affiliation{$^a$Laboratoire de Physique et M\'ecanique des Milieux H\'et\'erog\`enes (PMMH), CNRS, ESPCI Paris, PSL Research University, 10 rue Vauquelin, Paris, France; Sorbonne Universit\'e, Univ. Paris Cit\'e}
\affiliation{$^b$Faculty of Physics, University of Vienna, Kolingasse 14--16, 1090 Wien, Austria}
\affiliation{$^c$Institute for Theoretical Physics, TU Wien, Wiedner Hauptstr.\ 8--10, 1040 Wien, Austria}
\affiliation{$^d$Graduate School of Engineering Science, Osaka University, 5608531 Osaka, Japan}

\begin{abstract}
The migration of helical particles in viscous shear flows plays a crucial role in chiral particle sorting. Attaching a non-chiral head to a helical particle leads to a rheotactic torque inducing particle reorientation.  This phenomenon is responsible for bacterial rheotaxis observed for flagellated bacteria as Escherichia coli in shear flows. Here we use a high-resolution microprinting technique to fabricate micro-particles with controlled and tunable chiral shape consisting of a spherical head and helical tails of various pitch and handedness. By observing the fully time-resolved dynamics of these micro-particles in microfluidic channel flow, we gain valuable insights into chirality-induced orientation dynamics. 
Our experimental model system allows us to examine the effects of particle elongation, chirality, and head-heaviness for different flow rates on the orientation dynamics, while minimizing the influence of Brownian noise. Through our model experiments we demonstrate the existence of asymmetric bistability of the particle orientation perpendicular to the flow direction. We quantitatively explain the particle equilibrium orientations as a function of particle properties, initial conditions and flow rates, as well as the time-dependence of the reorientation dynamics through a theoretical model. The model parameters are determined using boundary element simulations and excellent agreement with experiments is obtained without any adjustable parameters. 
Our findings lead to a better understanding of chiral particle transport, bacterial rheotaxis and might allow the development of targeted delivery applications.

\end{abstract}


\maketitle




\section*{Introduction}
In nature, chirality occurs in a vast number of situations, often in the form of helical structures. At the nano and micron scale chirality can be encoded in the helical shape of relatively rigid objects, as for example DNA strands \cite{Watson1953}, cholesteric crystals \cite{Zastavker1999}, microorganisms such as spirochetes \cite{Nakamura2020} or bacteria flagella \cite{Lauga2016}. It also occurs dynamically, for example in the beating patterns of cilia in the lungs \cite{Gilpin2020}, the helical beating of the tails of sperm cells \cite{Gong2021, Cortese2021}, or of flagella of microalgae such as Chlamydomonas or Volvox \cite{Goldstein2015}. Natural and artificial microswimmers rely on the symmetry breaking induced by the chiral nature of flagella or microfabricated helices for self-propulsion at small scales \cite{Lauga2009a,Elgeti2015b, Tottori2012}.

Chirality also induces symmetry breaking in the transport properties of helical particles in interaction with viscous flows. Passive drift is observed for helical particles in shear flows \cite{Marcos2009, Makino2005,Meinhardt2012,Marichez2019,Ishimoto2020,Li2021c} in addition to the well-known Jeffery orbits of elongated objects. When a non-chiral head is added,  a rheotactic torque results, leading to a symmetry-breaking reorientation of the object \cite{Marcos2012,Jing2020} perpendicular to the flow direction. For a microswimmer these orientation dynamics lead to swimming into preferred directions so-called rheotaxis \cite{Marcos2012,Mathijssen2019,Jing2020}.

So far, experimental evidence of the above mentioned phenomena has exclusively \az{been} obtained using biological systems \cite{Marcos2009, Marcos2012, Mathijssen2019, Jing2020}. Despite their importance, their inherent complexity prevents isolating experimentally the role of specific particle properties on the observed phenomena. In addition Brownian noise often masks the effect of chirality \cite{Marcos2012, Jing2020}.

Recent developments in microfabrication methods give access to remarkable control of particle properties at the micrometer scale \cite{Stanton2015}. In particular 3D microprinting methods allow for the fabrication of well controlled rigid \cite{Stanton2015} or soft \cite{Wang2018} structures, that can be functionalized chemically  \cite{Ceylan2017} or through metal coatings \cite{Tottori2012, Baker2019, Xu2018}.

\begin{figure*}[tbh]
\centering
\includegraphics[width=\textwidth]{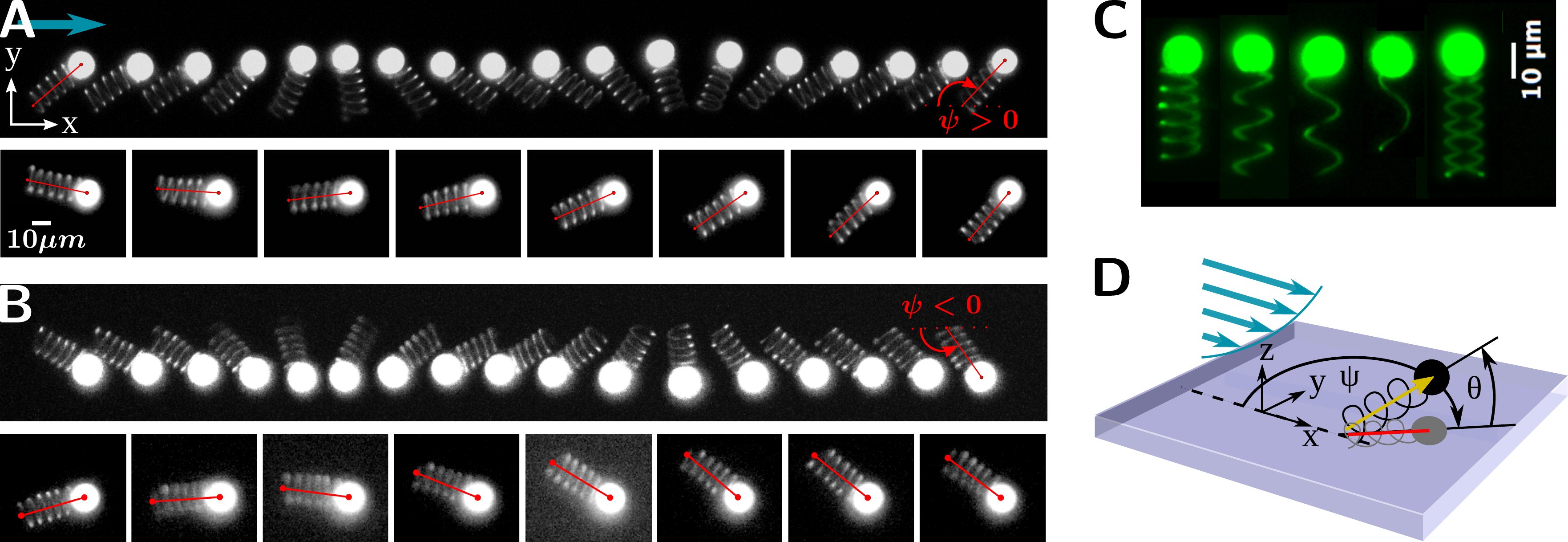}
  \caption{
    \az{(A,B)} Top: Subsequent snapshots \az{taken every $0.05~\mathrm{s}$} (from left to right)  in the $x$-$y$ plane of the orientation of a sphere with rigidly attached left-handed \az{(A) and right-handed (B)}  helix  transported to the right in microchannel flow. Bottom: snapshots of  maximum oscillation amplitudes, shown every 5 oscillation periods\azz{, i.e.\ taken every $\approx 3.5~\mathrm{s}$, } demonstrating the gradual reorientation perpendicular to the flow. 
      (\az{C}) Particles with different helix pitch $p=5,10,15,25~\mathrm{\mu m}$ (from left to right), and with two helices of opposite handedness with $p=10~\mathrm{\mu m}$.
      (\az{D}) Definition of coordinate system. 
      }
\label{fig:1}
\end{figure*}

Here we leverage these novel methods to overcome the limitations of biological systems and to design an experimental model system giving full control of particle shape, in a size range where Brownian noise can be neglected \cite{Zottl2019a}. We 3D print with sub-micron resolution rigid microparticles consisting of a microhelix attached to a slightly more dense spherical head.  They mimic the shape of flagellated bacteria without considering other complex properties \az{such} as activity while imitating  potential bottom-heaviness (here head-heaviness) of microorganisms \cite{Guasto2011, Pedley1992} or artificial micro-robots \cite{Tottori2012}. 

We investigate the orientation dynamics of these particles in viscous shear flows close to the bottom surface of a microfluidic channel. Direct observations under a microscope reveal for the first time the full time-resolved orientation dynamics of individual helical particles under flow (Fig. \ref{fig:1}\az{A,B}), in contrast to previous observations mostly limited to statistical information \citep{Marcos2009, Marcos2012, Jing2020}. We systematically vary the pitch of the microhelices (Fig.~\ref{fig:1}\az{C}) and as such the importance of chirality-induced reorientation effects. We develop a 
theoretical model which captures the experimental observations quantitatively without using any free fit parameters.

\section*{Particle design and fabrication} Chiral microparticles consisting of a spherical "head" and a helical "tail" are fabricated
  using a high resolution 3D microprinter (Nanoscribe) based on a 2-photon direct writing technique. Both spheres and helices have diameter $D=10~\mathrm{\mu m}$ and we produce helices of different pitch length $p=\{5,10,15,25\}~\mathrm{\mu m}$ with respective integer numbers of full turns $n=\{5,3,2,1 \}$, ensuring comparable particle aspect ratios, and resulting in helix lengths of $L=25-30~\mathrm{\mu m}$ (Fig.~\ref{fig:1}\az{C}).  Particles are fabricated with right- and left-handed chirality ($\chi=\pm1$).   Because of the production process the effective densities of the spheres ($\rho_s=1.263\pm0.015~\mathrm{g/cm^3}$) are higher than of the fabricated helices ($\rho_h\approx 1.20~\mathrm{g/cm^3}$). They are thus immersed in a background fluid made of polytungsten salt solution whose density has been approximately matched with the helix density, $\rho_f\approx \rho_h$ by adjusting the salt concentration (see \az{Materials and} Methods).

\section*{Microfluidic experiments} Experiments are performed in a shallow microfluidic channel with rectangular cross section of width $w=516~\mathrm{\mu m}$, height $h=96~\mathrm{\mu m}$ and length $2~\mathrm{cm}$, where a constant flow rate $Q$ is established by a syringe pump (Nemesys, Cetoni).
Particle observation occurs sufficiently far from the lateral walls to consider a quasi-planar Poiseuille flow in the \az{$x$}-direction $v_x=4v_c(h-z)z/h^2$, $z\in \{0, h\}$,   with a \az{$z$}-dependent shear rate $\dot{\gamma}=dv_x/dz$,   and  $v_c$ being the flow speed in the channel center \az{(see Fig.~\ref{fig:1}D for definition of coordinate system)}.

 After sedimentation to the bottom surface the \az{particle is oriented approximately parallel to the surface ($\theta_0 \approx 0$) and the}  initial in-plane ($x$-$y$) particle orientation $\psi_0$ (Fig.~\ref{fig:1}\az{D)}  can be tuned with a micromanipulator. Then the flow is switched on using flow rates $Q \in \{10,\dots,60\}~\mathrm{nl/s}$ and we follow the particle with a moving microscope stage, taking images with a Hamamatsu camera at 10 or 20 fps. The dynamics of the angle $\psi(t)$, $\psi \in \{-\pi,\pi\}$, is tracked by fitting particle orientations in the \az{$x$-$y$}-plane (red lines in Fig.~\ref{fig:1}\az{A,B,D}).  The out-of-plane angle \az{is denoted by}  $\theta \in \{-\pi/2,\pi/2 \}$ \az{(Fig.~\ref{fig:1}D). Its magnitude} can be estimated from the projected particle length, with a correction related to the particle thickness (see \az{Materials and} Methods).
 \az{In our images we cannot distinguish if the head points up or down, hence we cannot determine the sign of $\theta$.}

\section*{Experimental observations} 
\color{black}
Fig.~\ref{fig:1}A shows snapshots of typical dynamics of a left-handed ($\chi=-1$) microparticle under flow \az{at flow rate $Q=30~\mathrm{nl/s}$ for a particle of pitch $p=5~\mathrm{\mu m}$}.
\az{The first row depicts one full cycle of the observed} fast oscillation dynamics of $\psi(t)$, reminiscent of Jeffery orbits of elongated non-chiral neutrally buoyant particles \cite{Jeffery1922}. \az{Note that a Jeffery orbit corresponds to an oscillation in three dimensions and as such also the out-of-plane angle $\theta(t)$ undergoes periodic oscillations. These can be \azz{estimated} from the snapshots, but are less visible compared to the $\psi(t)$ dynamics that we will discuss in the following.}
\az{At the time scale of a single oscillation ($\sim 1~\mathrm{s}$) the sign of $\psi(t)$ does not change. Here $\psi(t)$ is always positive, with the head pointing in $+y$ direction, and the particle oscillates around $\psi=+\pi/2$ at a certain amplitude, as shown in the first row of Fig.~\ref{fig:1}A.
However, the orientation dynamics changes at longer time scales, as demonstrated in the} second row \az{for an initial orientation \al{$\psi_0<0$}. It}
shows the slow evolution of the orientation \az{angle $\psi$}, depicting a snapshot at maximum amplitude every 5 \az{oscillation} periods. 
\az{The particle starts oriented in $-y$ direction, 
then the orientation of the head switches ("flips")}
to the other side  ($\psi>0$) 
and \azz{eventually stabilize} at $\psi^\ast=+\pi/2$ 
(Movie S2).
\az{In contrast, when we perform the same set of experiments with right-handed particles ($\chi$=+1), flipping now occurs for initial orientations \al{$\psi_0>0$} which stabilize after flipping at \al{$\psi^\ast=-\pi/2$},
as demonstrated in Fig.~\ref{fig:1}B and Movie S5.}

\begin{figure*}
  \centerline{\includegraphics[width=\textwidth]{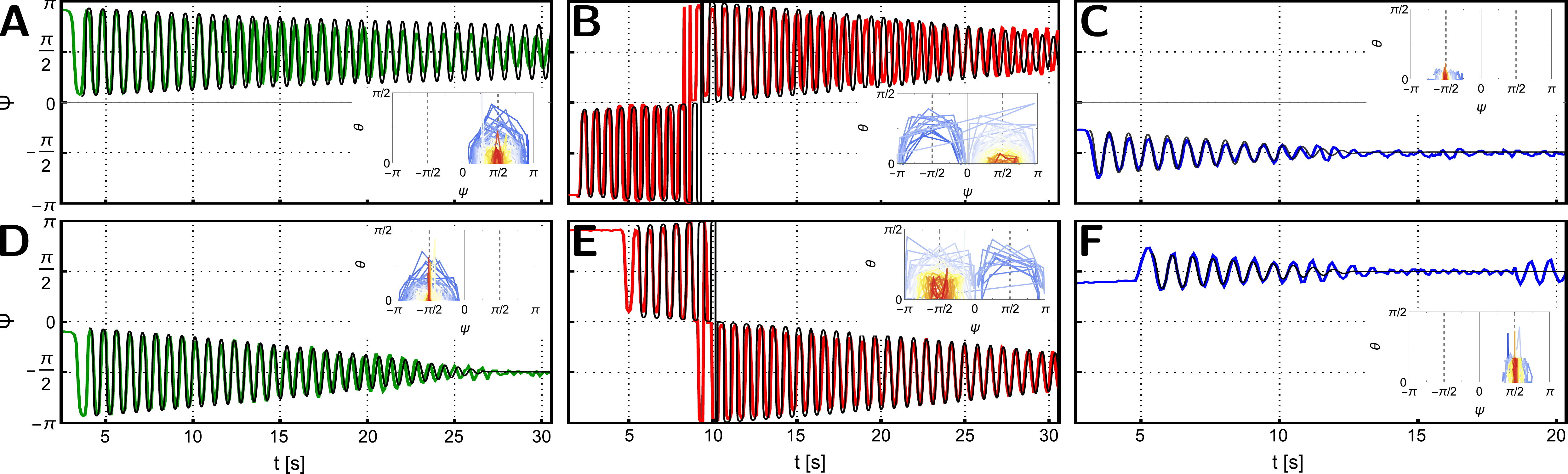}}
  \caption{Typical dynamics of orientation angle $\psi(t)$ for left- (A-C) and right-handed (D-F) particles with pitch $p=5~\mathrm{\mu m}$ for different initial conditions at flow rate $Q=30~\mathrm{nl/s}$ (colored curves) showing stabilization towards $\pm \pi/2$ (see movies S1-S6). The dynamics is quantitatively captured by a theoretical model [Eqs.~(\ref{Eq:dndt}) and (\ref{Eq:dndt2}), (black curves)].  Model shear rates have been adjusted to match experimental oscillation frequencies and are in the range $\dot{\gamma}=27\pm 1.5~\mathrm{s^{-1}}$, within $\pm 5\%$ from independently determined experimental shear rates.  Insets: experimental phase space trajectories ($\psi,|\theta|$) with time color-coded (from blue to red).
  }
\label{fig:2}
\end{figure*}

\az{
\al{Fig.~\ref{fig:2} depicts quantitative measurements of $\psi(t)$ for left- and right-handed \azz{particles} for the same conditions as \azz{in} Fig.~\ref{fig:1} ($Q=30~\mathrm{nl/s}$, $p=5~\mathrm{\mu m}$) now varying the initial orientations $\psi_0$.} \az{Different dynamics can be observed for \al{identical handedness} as a function of the initial orientation, as shown in the \al{top row} for left-handed particles ($\chi=-1$). When} the initial $\psi_0$ is \al{positive, the particle oscillates around $\pi/2$ and the oscillation amplitude decreases with time until the particle stabilizes at $\psi^\ast = +\pi/2$ (Fig.~\ref{fig:2}A and Movie S1). When the initial  $\psi_0$ is negative and the particle oscillates with a sufficiently large amplitude, 
an increase in amplitude is observed until the particle flips to the other side and stabilizes again at $\psi^\ast = +\pi/2$ (Fig.~\ref{fig:2}B and Movie \az{S2}) similarly to what is shown in Fig.~\ref{fig:1}A. For} \az{negative $\psi_0$} and sufficiently small oscillation amplitudes 
however, the particle does not flip to the other side, but the oscillation amplitude decreases with time and the particle is stabilized at $\psi^\ast = -\pi/2$\az{, as shown in} Fig.~\ref{fig:2}C and Movie S3.
We thus observe bistability with final particle orientation at $\psi^\ast_1 = \pi/2$ or $\psi^\ast_2 = - \pi/2$\az{, depending on the initial orientation $\psi_0$}. This bistability is asymmetric, as particles with $\psi_0<0$ can flip to the other side, whereas particles with $\psi_0>0$ never flip but are always stabilized at $\psi^\ast_1 = \pi/2$. \al{We denote the equilibrium position that can also be reached through flipping \azz{by} $\psi^\ast_1$ and the other equilibrium position \azz{by} $\psi^\ast_2$. The dynamics are classified as stabilizing at the more stable orientation $\psi_1^\ast$ without flipping (green) or with flipping (red) and stabilizing at the less stable orientation $\psi_2^\ast$ (blue).} 
}

\az{To demonstrate that this asymmetry stems from the particle chirality, we perform experiments 
also for right-handed particles ($\chi$=+1). Indeed,}
we observe exactly the opposite dynamics.
\az{While right-handed particles again stabilize either at $\psi^\ast = \pi/2$ (Fig.~\ref{fig:2}F and Movie S6)  or at $\psi^\ast = - \pi/2$ (Fig.~\ref{fig:2}D,E and Movies S4 and S5),  flipping now only occurs for initial orientations $\psi_0> 0$ (Fig.~\ref{fig:2}E).} \al{We can thus write the equilibrium positions in a compact form as $\psi^\ast_1= -\chi \pi/2$ and $\psi^\ast_2= +\chi \pi/2$.}

\al{A more direct comparison with classical Jeffery dynamics becomes possible when considering the angular phase space dynamics as shown in }the insets of Fig.~\ref{fig:2}. \al{The experimentally measured dynamics ($\psi(t),|\theta(t)|$) reveal the coupling between in-plane and out-of-plane oscillations. Large amplitudes in $\psi$ also correspond to large amplitudes in $\theta$, with the maximum amplitude in $\psi$ for \azz{$\theta=0$,} corresponding to particles aligned with the surface, and the maximum amplitudes for $\theta$ to $\psi=\pm \pi/2$. A decrease in amplitude of $\psi$ and a stabilization towards \az{$\pm \pi/2$} also corresponds to a decrease in amplitude of $\theta$ \az{towards zero} and thus particle alignment with the surface. At short times the dynamics reiterates Jeffery-like oscillatory dynamics, however damping and flips are observed at longer times. 
This is evidently in contrast to classical Jeffery orbits where oscillation amplitudes are constant and $\psi(t)$ is either positive or negative for the entire length of the trajectory \cite{KimKarila}.}


\az{We note that in some cases, after the particle orientation reaches a stable orientation, it assumes a "kick" which re-starts the oscillation process at small amplitude, as depicted in Fig.~\ref{fig:2}F at starting time $t \approx 18s$. We attribute such disruptions to imperfections, such as small impurities  in the channel wall.} 
 
\color{black}
\section*{Model} To understand the observed particle dynamics, we develop a 
theoretical model, assuming a constant shear rate $\dot{\gamma}$ experienced by the particles moving close to the bottom wall, and  neglecting hydrodynamic particle-wall interactions and the $z$-dependent shear rate \cite{Pozrikidis2005} (see  discussion below\az{)}. Then the dynamic equations for \az{the orientation angles} $\psi$ and $\theta$  can be written as
\begin{align}
  \frac{d \psi}{dt} =&  \dot{\gamma}\left(1+\alpha^{-2}\right)^{-1}\sin\psi\tan\theta 
  - \chi \dot{\gamma} \nu \cos\psi \cos 2 \theta/ \cos\theta 
  \label{Eq:dndt} \\
  \frac{d \theta}{dt} =&
\frac{1}{2}\dot{\gamma}(1 - (\alpha^2-1)(\alpha^2+1)^{-1}) \cos 2\theta)\cos\psi  \nonumber \\
                       &  - \chi \dot{\gamma} \nu \sin\psi\sin\theta
   -\Omega_H \cos\theta \mathcal{H}(\theta)
    \label{Eq:dndt2}
    \end{align}
  including three contributions:  The first terms in Eqs.~(\ref{Eq:dndt}) and (\ref{Eq:dndt2}) describe Jeffery oscillation dynamics of elongated particles with \az{effective} aspect ratio $\alpha$ 
  in shear flows \cite{Jeffery1922}.  The second terms describe chirality-induced reorientation  of particles of handedness $\chi$ and \az{dimensionless} "chiral strength" $\nu$ depending on the shape of the helix and the size of the spherical head \cite{Mathijssen2019,Jing2020}. 
  \az{This term rotates particles consisting of a non-chiral head and a chiral tail, such as bacteria, towards the positive or negative vorticity direction of the flow, depending on the chirality of the tail \cite{Mathijssen2019,Jing2020}. The strength of the chiral reorientation rate is a product of the shear rate and chiral strength, $\Omega_C=\dot{\gamma}\nu$. Indeed,}
  it has previously been shown that this \az{chirality-induced reorientation} leads to lateral drift of  swimming bacteria in shear flows although direct experimental validation on individual trajectories has not been achieved yet  \cite{Marcos2012,Mathijssen2019,Jing2020}.
  The third term in Eq.~(\ref{Eq:dndt2}) reflects that the head of the particle is heavier than the helical tail and is the only term not proportional to $\dot{\gamma}$. 
  This "head-heavy" torque  rotates particles towards head-down orientations. In our simplified model, it only depends on  $\theta$, and a constant head-heavy strength $\Omega_{H}$ which depends on particle shape and linearly on the density difference $\Delta\rho=\rho_s-\rho_f$.  It vanishes when the particle head (almost) touches the wall hence sedimentation is suppressed.
  This we capture roughly with the Heaviside function $\mathcal{H}(\theta)=1$ for $\theta>0$ and 0 otherwise, meaning that this torque is suppressed when the particle points towards the wall ($\theta<0$), see also  SI Appendix\az{, SI Text and Fig.~S4}.

\begin{figure}
\centering
\includegraphics[width=\linewidth]{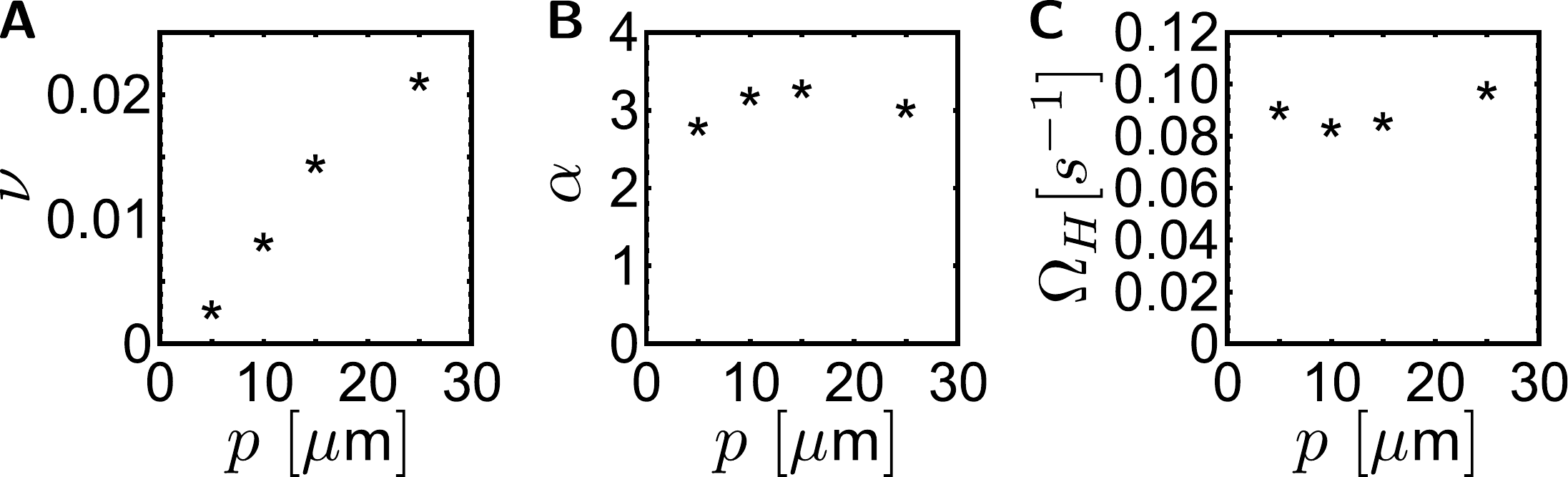}
\caption{Parameters obtained from BEM simulations for particles of different pitch $p$. (A) Chiral strength $\nu$. (B) Effective particle aspect ratio $\alpha$. (C) Head-heavy reorientation strength $\Omega_H$. 
}
\label{fig:4}
\end{figure}

To determine numerical values for the particle properties  $\alpha$, $\nu$ and $\Omega_H$ we use the boundary element method (BEM) with a triangulated mesh for the surface of the particles immersed in Stokes flow (see \az{Materials and} Methods)\cite{Pozrikidis1992,Ishikawa2006,Matsunaga2017}.
We perform two independent sets of BEM simulations.
First we determine the particle aspect ratios  $\alpha$ and the chiral strengths $\nu$ for all the different experimental chiral particle shapes by considering  neutrally buoyant right-handed particles ($\Omega_H=0$, $\chi=+1$) in simple shear flow.
To determine $\nu$ and $\alpha$ we place
the particle aligned with the flow ($\psi = 0$, $\theta=0$) at a given shear rate and measure the instantaneous angular velocities for different orientations along the particle axis, i.e. different phase angles where the helix is anchored to the head. By averaging over these orientations  we can immediately determine $\nu$ and $\alpha$ from Eqs.~(\ref{Eq:dndt}) and (\ref{Eq:dndt2}).

\begin{figure*}[tb]
  \centerline{\includegraphics[width=15cm]{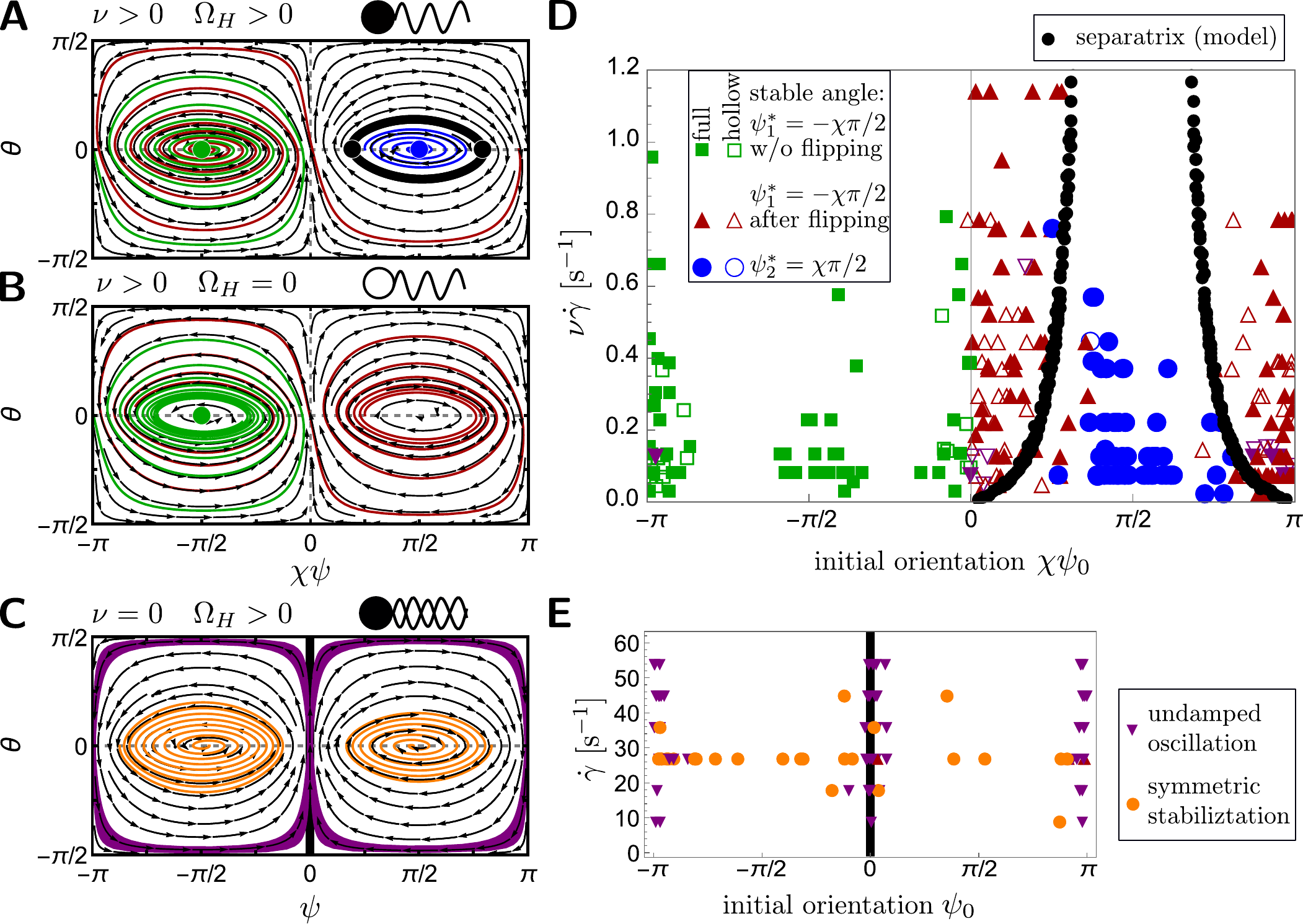}}
  \caption{(A,B) Orientation \az{phase space}  from \az{the theoretical} model\az{, Eqs.~(\ref{Eq:dndt}) and (\ref{Eq:dndt2}),} for right- ($\chi=+1$) and left-handed ($\chi=-1$) particles shown for chiral strength $\nu=0.0212$ and shear rate $\dot{\gamma}=9~\mathrm{s^{-1}}$ for head-heavy (A) and neutrally buoyant (B) particles. 
  Initial orientation $\chi\psi_0<0$ (green \az{trajectories}) leads to stabilization at \azz{$\chi \psi_1^\ast =-\pi/2$ (green dot), i.e.\ $\psi_1^\ast =-\chi\pi/2$,} while initial $\chi\psi_0>0$ can either lead to stabilization at \azz{$\psi_1^\ast=-\chi\pi/2$} (red flipping trajectories), or at \azz{$\psi_2^\ast=\chi\pi/2$} (blue), see also Fig.~\ref{fig:2}.
  (\az{D}) Experimentally observed asymmetric bistability for different  $\dot{\gamma}$,  $\nu$ and  $\chi$ for head-heavy (filled symbols) and hollow-head (empty symbols) particles depending on initial orientation $\chi\psi_0$:
   Stabilization at \azz{$\psi_1^\ast=-\chi\pi/2$} (green for initial $\chi\psi_0<0$ and red for $\chi\psi_0>0$) and at \azz{$\psi_2^\ast=\chi\pi/2$} (blue, occurs only for $\chi\psi_0>0$).  
  (\az{C},E) Characterization of non-chiral double-helix particles from \az{theoretical} model (\az{C}) and experiment (E) which are either stabilized on the initial side (orange), or perform Jeffery-like tumbling trajectories not approaching a stable point (purple).
 Thick black curves in (A,\az{C},E): Separatrix from \az{theoretical} model dividing the stable regions. \az{Black dots in (A,D): Separatrix values at $\theta_0=0$. 
 }
    }
\label{fig:3}
\end{figure*}

  Second, $\Omega_{H}$ is determined by a different set of BEM simulations. We put a particle with a heavy head of \az{the experimentally measured} density $\rho_s$ which is initially
  aligned perpendicular to the direction of gravity ($\theta_0=\pi / 2$) in a quiescent fluid \az{(}$\dot{\gamma}=0$\az{)} of \az{the experimentally measured} density $\rho_f$. We then measure the angular velocity due to head-heaviness, and directly obtain $\Omega_H$ from Eq.~(\ref{Eq:dndt2}).
  
  The values of $\nu$, $\alpha$ and $\Omega_H$ obtained from the BEM simulations for  particles of different pitch $p$ are plotted in Fig.~\ref{fig:4}. 
   $\Omega_H$ and $\alpha$ are comparable for all helix shapes, i.e.\  $\alpha\in \{ 2.8, 3.2 \}$ and $\Omega_H \in \az{\{ 0.084, 0.098 \}}~\mathrm{s^{-1}}$. Note, $\alpha$ is somewhat smaller than expected from naive estimates $\alpha_0=(L+D)/D \in \{3.5,4\}$. In contrast, $\nu$ increases significantly with helix pitch $p$ from $\nu=0.003$ ($p=5~\mathrm{\mu m}$) to $\nu=0.02$ ($p=25~\mathrm{\mu m}$).
  Hence,  tuning the pitch allows to adjust the chiral strength of the particle, while leaving effective aspect ratio and head-heaviness approximately unaffected.

The parameters $\nu$ and $\alpha$ are defined and determined in simple shear flow, while our experimental Poiseuille flow is characterized by non-constant and wall-bounded shear. To verify the simple shear approximation we also determine effective values of $\nu$ and $\alpha$ in Poiseuille flow, both with and without the presence of bounding walls using BEM. Indeed we show that the effect of the quadratic Poiseuille flow profile  on  $\nu$ and $\alpha$  is very small.
The effect of hydrodynamic interactions with the wall slows down the Jeffery-like reorientation close to the wall, as expected \cite{Pozrikidis2005}, but only has a small effect on the chiral reorientation (see SI Appendix\az{, SI Text and Fig.~S6}).

 The shear rates $\dot{\gamma}$ experienced by oscillating particles are estimated by two independent methods \az{from the experimental data}.
 First,  
 from the experimentally measured particle velocities $v_x(t)$ at a given flow rate $Q$
 we can estimate the particle position $z$ and eventually the local shear rate.
 Second,  we determine the oscillation frequencies from the maxima of the \az{power spectrum}
 of \az{the experimental orientation dynamics} $\psi(t)$.
 The results from both methods consistently show that $\dot{\gamma}$ can be calculated from the flow rate $Q$ as $\dot{\gamma}\approx 0.9 Q~\mathrm{nl^{-1}}$   (see Materials and Methods).

 All together we end up with a theoretical model, Eq.~(\ref{Eq:dndt}) and (\ref{Eq:dndt2}), without free parameters and which can, after numerical integration to obtain $\psi(t)$ and $\theta(t)$, be directly compared to the experiments.
The model reproduces the experimental trajectories $\psi(t)$ extremely well, see Fig.~\ref{fig:2} (black curves), including oscillation frequencies, amplitude modulations, flipping behavior and stable positions.

\section*{Discussion} 
\color{black}
\al{Combining experimental and theoretical results we can now analyze the different particle dynamics and their origins in detail.} Fig.~\ref{fig:3}A depicts trajectories in  \az{orientation} phase space from the theoretical model  \az{(Eq.~(\ref{Eq:dndt}) and (\ref{Eq:dndt2}))}, where \al{$\theta$ is represented as a function of $\chi \psi$ to collapse the results for different handedness onto one graph}. The different dynamics (stabilizing at $\psi_1^\ast$ without and with flipping (red and green) and stabilizing at \azz{$\psi_2^\ast$} (blue))
are indicated using the same color code as in Fig.~\ref{fig:2}. The results agree \al{qualitatively} with the experimental observations shown on the insets of Fig.~\ref{fig:2}.  
Reaching either of the two stable orientations depends on the initial condition, and a separatrix \az{(black curve)}  divides the two stable regions in phase space. \al{We will demonstrate below that} 
\az{the separatrix and the size of the stable regions depend on the ratio of $\Omega_C$ $(=\nu\dot{\gamma})$ and $\Omega_H$. \al{For constant $\Omega_H$}, the larger $\nu\dot{\gamma}$, the smaller is the region of initial $\psi_0$ values which approach the less stable position $\psi_2^\ast$. }

In Fig.~\ref{fig:3}\az{D} we summarize the experimental results for different chiral strength $\nu$, handedness $\chi$ and shear rate $\dot{\gamma}$ as a function of \az{initial condition} $\chi \psi_0$ and the 
product $\nu\dot{\gamma}$. \az{Again, we classify the results as a function of the three types of trajectories
using the same color code as in Fig.~\ref{fig:2} and in Fig.\ref{fig:3}A}. Here $\Omega_H$ is kept approximately constant, while $\dot{\gamma}$ and $\nu$ are modified independently.

\az{Since in the experiments the initial angle $\theta_0 \approx 0$, solely the value of the initial angle $\psi_0$ determines the final stable position, bounded by the two $\psi$-values of the separatrix at $\theta=0$ (black dots in Fig.~\ref{fig:3}A).
Fig.~\ref{fig:3}D shows these separatrix values as black dots
for many different combinations of $\nu$ and $\dot{\gamma}$ values and for constant $\Omega_H=0.09~\mathrm{s^{-1}}$.
Indeed they only depend on the product $\nu\dot{\gamma}$, and the region to reach $\psi_2^\ast$ decreases with increasing  $\nu\dot{\gamma}$.}

\az{We observe that the theoretically predicted separatrix}
is in good agreement with experimental results, 
and our findings unambiguously demonstrate an asymmetric, handedness-dependent bistability of chiral head-heavy particles in flow.

\al{To further show the robustness of our model and the necessity of reorientation contributions from both head-heaviness and particle chirality to observe asymmetric bistability,  we experimentally and numerically "knock out" each of these contributions separately. }

\az{When head-heaviness is ignored in the simulations ($\Omega_H=0$), Fig.~\ref{fig:3}B, the chiral reorientation rate $\Omega_C$ alone determines the rise and decay of the amplitudes.} \al{All particles will stabilize at $\psi_1^\ast=-\chi \pi/2$, including flipping of those with initial orientation of opposite sign and where the sign of $\psi_1^\ast$ is solely given by the handedness.}
\az{Thus no bistability is observed and no separatrix exists in Fig.~\ref{fig:3}B.}
In the experiments we create particles with hollow spherical heads having a density comparable to the helix density, strongly reducing head-heaviness (see \az{Materials and} Methods). Then almost all particles end up pointing to \az{$\psi_1^\ast=-\chi\pi/2$}, see Fig.~\ref{fig:3}\az{D} (open symbols)\az{, as expected from the theoretical model.}

On the other hand when head heaviness is considered in the absence of chirality ($\nu=0$) in the simulations two different types of dynamics are observed (Fig.~\ref{fig:3}C):
Oscillations either decay symmetrically to $\psi^\ast=\pm \pi/2$ (orange), or they oscillate practically undamped (violet) when particles start at large amplitude i.e.\ very close to the separatrix (black line at $\psi=0$). The decay \az{towards symmetric} equilibrium positions correspond to bistability separated by a separatrix at $\psi=0$ (black line). 
\az{In the experiments} we produce particles with two helices, with opposite handedness but same pitch  (Fig.~\ref{fig:1}\az{C}, right).
\az{For these particles the two opposite handedness cancel out the two opposite chiralities leading to effectively non-chiral particles with $\nu\approx 0$.}
\az{For the double-helix particles we indeed do not observe asymmetric bistability, but obtain the two different types of dynamics shown in Fig.~\ref{fig:3}E.}


\az{Hence, we have shown from experiments and modeling that the combination of chirality and head-heaviness leads to the asymmetric bistability shown in Fig.~\ref{fig:3}A and D. Reorientation due to head-heaviness is responsible for bistability and reorientation due to chirality for the asymmetry. 
The asymmetry of the $\psi$-$\theta$ phase-space for the two different handedness ($\chi=\pm 1)$ can be captured by the same phase space when plotted in an $\chi\psi$-$\theta$ phase space, as in Figs.~\ref{fig:3}A,D. For a random initial condition it is more likely to end up at the stable position $\psi_1^\ast=-\chi \pi/2$, which can also be reached through flipping (red trajectories in Fig.~\ref{fig:3}A), compared to the less probable stable position $\psi_2^\ast=\chi \pi/2$.} 

\begin{figure*}[bt]
\centering
\includegraphics[width=11.4cm,height=11.4cm]{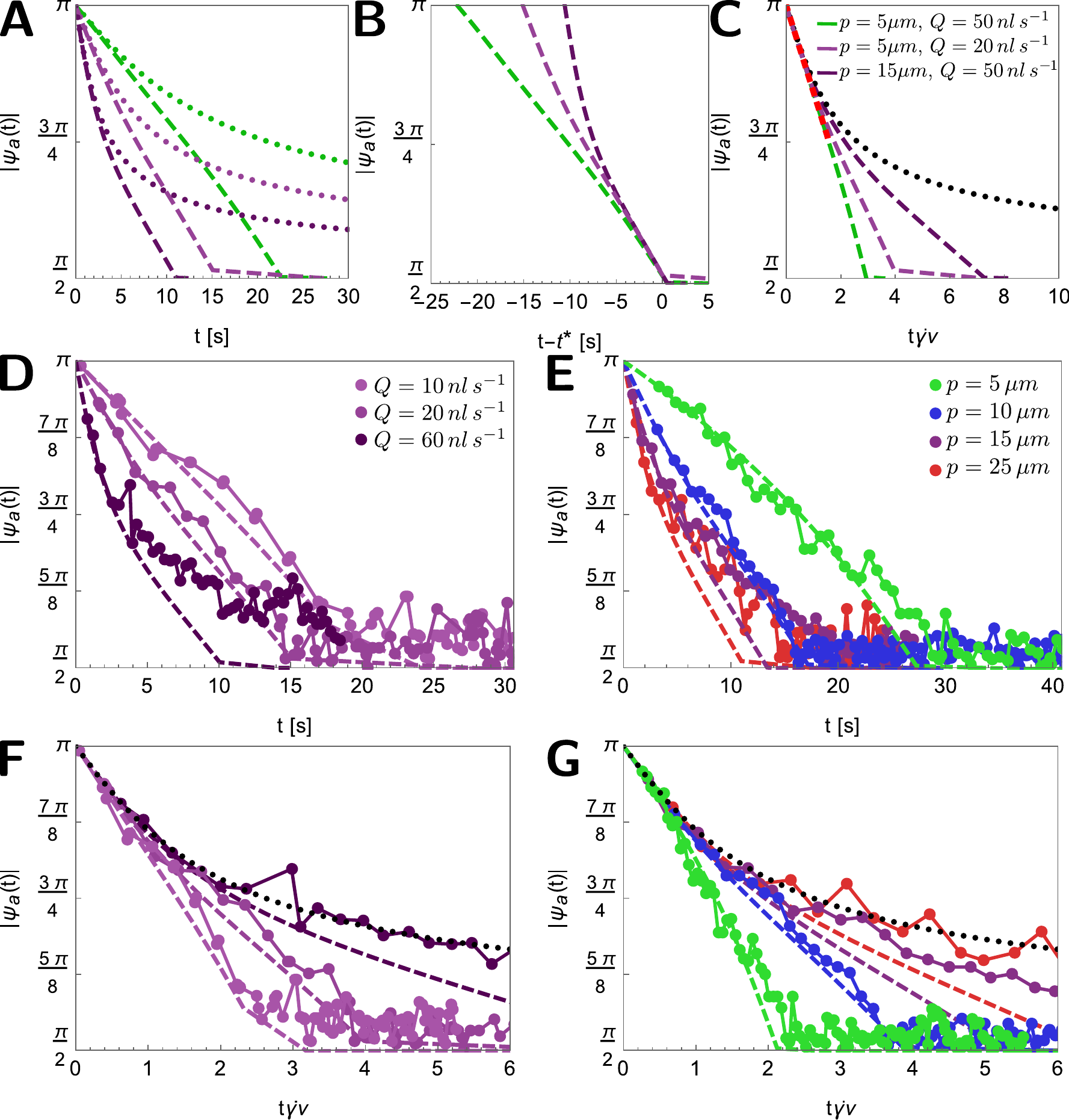}
  \caption{Oscillation amplitude decay $|\psi_a(t)|$ for chiral head-heavy particles.
  (A-C) $|\psi_a(t)|$ \az{obtained from the theoretical model (Eqs.~(\ref{Eq:dndt}) and~(\ref{Eq:dndt2}))} for three  combinations of pitch $p$ (i.e.\ $\nu$) and flow rate $Q$ (i.e.\  $\dot{\gamma}$) depending on time \az{$t$} (A), shifted time \az{$t-t^\ast$} (B), and rescaled time \az{$t\dot{\gamma}\nu$ (C), where $t^\ast$ is the time $|\psi_a(t)|$ has decayed to the stable orientation $\pi/2$. Dashed lines show results from the full model, dotted lines in (A) for non head-heavy particles ($\Omega_H=0$).} 
  (D) $|\psi_a(t)|$ \az{from typical experimental trajectories (dots) and from the model (dashed lines)} for pitch $p=15~\mathrm{\mu m}$ for different  $Q$.    (E) $|\psi_a(t)|$ for  $Q=30~\mathrm{nl/s}$ for different $p$.
    (F,G) Respective time-rescaled plots of (D) and (E).
    \az{The} black dotted curves in (C,F,G) show \az{the} universal \az{parameter-independent} decay \az{of $|\psi_a(t)|$} for neutrally buoyant particles \az{when plotted against rescaled time  $t\dot{\gamma}\nu$. The }initial decay $|\psi_a(t)| \approx \pi-t\dot{\gamma}\nu/2$ \az{is indicated by the} red dotted line in (C).
  }
\label{fig:5}
\end{figure*}

\color{black}
\section*{Decay rates}
The particle geometry also determines the timescales to reach the final stable orientations. Fig.~\ref{fig:5} shows the decay of the oscillation amplitudes $|\psi_a|$ (after potential flip) from $\pi$ towards the stable orientation at $\pi/2$ for different flow rate $Q$ (or $\dot{\gamma}$) and particle pitch $p$ (or $\nu$). Results from our model are shown in Fig.~\ref{fig:5}A (dashed curves). Starting from large amplitudes, the initial decay is faster for larger $\dot{\gamma}$ and stronger $\nu$, meaning larger $p$ and $Q$, influencing the times $t^\ast$ when $|\psi_a(t)|$ has decayed to the stable value $\pi/2$. For smaller angles the decay is dominated by head-heaviness  and is independent of $\nu$ and $\dot{\gamma}$. 
Indeed in Fig.~\ref{fig:5}B, when plotted against $t-t^\ast$, constant slopes can be identified.
Removing the effect of head-heaviness (\az{$\Omega_H=0$,} dotted curves \az{in Fig.~\ref{fig:5}A}) does not influence the initial decay but  slows down the approach \az{to} the stable position (see also Fig.~\ref{fig:3}\az{B}). For large amplitudes the decay is fully given by chirality-induced reorientation  and the amplitude decay collapses when plotted against the rescaled time $t\dot{\gamma}\nu$, see Fig.~\ref{fig:5}C, with $|\psi_a(t)|\approx \pi-t\dot{\gamma}\nu/2$ (red dotted line).  In the absence of head-heaviness the amplitude decay collapses to a universal curve (black dotted curve in Fig.~\ref{fig:5}C\az{)}. 
We obtain quantitative agreement between  model and experiments in particular  for large amplitudes
\az{This is} demonstrated in Fig.~\ref{fig:5}D \az{for different flow rates $Q$ and fixed particle shape (pitch $p=15~\mathrm{\mu m}$), and in Fig.~\ref{fig:5}E for different particle pitch $p$ at fixed flow rate $Q=30~\mathrm{nl/s}$. 
Here we have plotted  the experimental amplitudes along with the theoretical curves (dashed lines), in very good quantitative agreement.
Fig.~\ref{fig:5}F,G show the respective plots using rescaled time $t\dot{\gamma}\nu$, and indeed the experimental data collapses well for large amplitudes, in agreement with the theoretical model.}
\az{As can be seen in Fig.~\ref{fig:5}F,G, }experimental results sometimes start to deviate towards slower decays at small angles, and even approach the limit of no \al{head}-heaviness (black dotted lines). We attribute this to a reduction of the importance of head-heaviness due to fluctuating decrease of the experimental density difference. This is particularly pronounced for large $\dot{\gamma}$ or $\nu$, where chirality-induced reorientation is strong (see \az{also SI Appendix, Fig.~S7, Fig.~S8 and Fig.~S9}.

\section*{Conclusions}
Through a combination of highly resolved experiments and a theoretical model we have demonstrated novel asymmetric bistable orientation dynamics of chiral microparticles in shear flows. The interplay between particle elongation, chirality and head-heaviness is fully captured by an analytical model without adjustable parameters, independently determined from BEM simulations. Our results constitute the first direct experimental observation and quantitative comparison of individual helical particle orientation dynamics. \az{The findings of our work will be helpful} to
better understand dynamics in more complex biological systems and to design artificial microrobots or targeted delivery applications.  

\section*{Materials and Methods}
\subsection*{Particle Printing}
The particles are fabricated with a two-photon lithography micro-printer by Nanoscribe, using the Dip-in operational mode with the IP-Dip photoresist (Nanoscribe). The helix is printed at the smallest available resolution, with its cross-section given by the convolution of one voxel (grain of rice shape with $1~\mathrm{\mu m}$ in height and less than $140~\mathrm{nm}$ in diameter)  and helix radius $5~\mathrm{\mu m}$.
The spherical body is also printed with the same radius, either as a full sphere by multiple layers (slicing distance set to $50~\mathrm{nm}$ to smooth the step between the layers), or as a spherical shell of $1~\mathrm{\mu m}$ thickness. In the case of the shell, the sphere is open on the side of the helix with a circular cut with radius $2~\mathrm{\mu m}$, in this way the unpolymerized resin can be washed away during the development and it can be substituted with the desired fluid. The particles are printed on an array configuration with the axis of the helix parallel to the quartz glass substrate (see \az{SI Appendix,}  Fig.~S1), 
and at a small vertical distance from the substrate so that the helix cross section is fully resolved (distance of the centerline of the helix from the substrate between $0.4$ and $0.7~\mathrm{\mu m}$). The sample is left at rest for at least 24 hours after the printing, in order for it to gain stiffness and better sustain the development processes. For developing the particles, the sample is put 20-30 minutes in PGMEA (Propylene Glycol Methyl Ether Acetate, from Sigma).
To avoid capillary forces which could deform and destroy the helices, the sample is then put for a few seconds into distilled water, then 2 minutes in Isopropanol, and again in water.
The outcome particle can be considered rigid under our experimental flow conditions. 
After the development, the sample is stored in water with a small amount of Sodium hypochlorite, and BSA at concentration $0.1\%$, to avoid bacteria growth and particles stickiness. 

\subsection*{Particle density}
To measure the density of the head we
print the spherical body alone, full or hollow and without helix, to perform sedimentation tests.
The individual spherical particle (or spherical shell) is let sinking inside a PDMS chamber in $50\%$ glycerol with no flow, and its position along the sedimentation axis is recorded by continuously adjusting it relative to the focal plane, using a motorized stage, for a distance of around $300~\mathrm{\mu m}$. The sphere settling velocity is then used to calculate the particle density  by balancing the gravitational force and the buoyancy with the drag given by the Stokes' law. The measured density of the polymerized material, estimated by the sedimentation of printed spheres, appears to fall in the interval known in literature, which is between $1.190$ and $1.370~\mathrm{g/cm^3}$ \citep{Bauer2014}, but it appears to be also dependent on the exact printing procedure. We measure for the full sphere a density of $1.26~\mathrm{g/cm^3}$, while the density of the helix alone is systematically smaller (estimated to be around $1.20$g/cm$^3$ by repeated sedimentation experiments at increasing fluid density). This unavoidable density inhomogeneity is also confirmed by the fact that the whole particle given by a full sphere and the helix in bulk $50\%$ glycerol is observed falling parallel to gravity with the helix pointing upwards. The hollow sphere geometry, instead, provides a better mass distribution, with an average density of the sphere typically from 3$\%$ to 6$\%$ smaller than the full printed version, much closer to the one of the helix alone.

\subsection*{Microfluidic device}\label{appA2}
The experimental channel consists of a rectangular PDMS shallow channel, build with standard soft-lithographic techniques, with nominal dimensions $500~\mathrm{\mu m}$ in width and $100~\mathrm{\mu m}$ in height (measured $516~\mathrm{\mu m} \times 96~\mathrm{\mu m}$). 
The channel is connected from one side to a syringe pump while the other side is in direct communication to a large pool where the sample with the particles is located. The pool is open on the surface in a way that the experimenter has access to the sample from above and the particles can be individually manipulated and transported to the entrance of the channel. The manipulation is performed using a thin capillary, with nozzle size $\lesssim 8~\mathrm{\mu m}$, whose position is controlled by a micromanipulator (Eppendorf) connected to a $3$ mL plastic syringe. A flow can be manually imposed through the nozzle in both directions and it is used for grabbing the particle from the sample, by imposing a negative pressure, and releasing it at the entrance of the channel, with positive pressure. The capillary nozzle is functionalized with $2\%$ BSA for 30 minutes before use in order to avoid other non-specific capillary-particle interactions.
The PDMS channel is sealed to a cover glass coated with a thin layer of PDMS in a way that PDMS also covers the bottom wall, which is in contact with the particle during the experiment. The channel is also treated before use with $3\%$ BSA. This facilitates the experiment with the particle in the vicinity of the wall avoiding any wall-particle interaction apart from steric interaction.

The particle is usually placed at the entrance of the channel, in the vicinity of the bottom wall and sufficiently far away from the lateral walls. 
The device is placed under a microscope (Zeiss Axio Observer), on a MS-2000 automated stage (ASI), which allows to follow the particle along the whole length of the channel. The capillary has access to the entrance of the channel on the side of the pool and it can be also used to change, to some extent, the particle initial orientation $\psi_0$.

\subsection*{The fluid and flow control}

We use polytungstate solution (PTS) at concentration of $2.7$g of salt dissolved in $10~\mathrm{mL}$ of water to match approximately the density of the helical structure, while the spherical head remains typically denser. The corresponding viscosity is $\eta=1.17~\mathrm{mPas}$. At this concentration evaporation from the open pool induces variations in the fluid density and viscosity that has been measured to be limited below $5\%$. 
The device is placed inside a transparent box which is open on one side for the particle manipulation, to minimize the solvent loss.
The flow is provided by a Hamilton glass $500~\mathrm{\mu L}$ syringe and controlled by a Nemesys syringe pump.
The flow is switch\az{ed} on after the particle has deposited to the bottom wall of the channel and it is switched off before the particle exit\az{s} the channel. A typical experiment use the same particles over many runs, by reversing the flow several times.

\subsection*{Image acquisition and image processing}

The particle in the channel is visualized through a 20X long working distance objective with fluorescence microscopy using a high efficiency filter (BP 430/60, BP 550/100 Zeiss) and images are taken with a Hamamatsu Orca-flash 4.0LT camera, working at 10 or 20 Hz and synchronized with the dumping of the stage position, so that also the particle translation can be completely reconstructed. 
After a median filtering, the location of the center of mass of the sphere is located either by setting a threshold for the case of the full sphere or by circle detection for low intensity spherical shells. Then the treatment proceeds at the level of the helix: its shape, especially at the tip, is reconstructed by a maximum filter and the image is binarized. The helix orientation is found by considering the maximum overlap between the helix and a rotating rectangle pinned at the center of the sphere and with width similar to the helix. The helix projected length $L_p$ is also extrapolated fixing a threshold on the profile of integrated intensity over the rectangle along the correct orientation and rescaling this value by using a measurement of a known projected length.

The absolute value of the out-of-plane angle $|\theta|$ can then be estimated from the projected length $L_p$, the helix length $L$, and the helix/sphere diameter $D$, such that $|\theta| \approx \arccos((L_p-D)/L)$.

\subsection*{Estimation of local shear rate}
\label{sec:shear}
First, $\dot{\gamma}$ has been determined from the experimentally measured particle velocities $v_x(t)$.
When we neglect hydrodynamic particle-wall interactions and the flow curvature, we can assume to first approximation that the particles simply follow the flow velocity $v_x(y,z)$ in the rectangular channel.
For a given channel geometry and applied flow rate $Q$ the flow field $v_x(y,z)$ in the channel can be calculated, with $v_\text{max}$ the velocity in the center of the channel \cite{Bruus2007}.
Since our particles are sufficiently far away from the side walls, the flow profile can be approximated as planar Poiseuille flow, $v_x(z)=(4/h^2)v_{max} (h-z)z$ with $h=96~\mathrm{\mu m}$.
Measuring the particle velocity and comparing to the flow profile can then be used to extract the particle position $z$ in the channel, and eventually the $z$-dependent shear rates $\dot{\gamma}(z)=dv_x/dz$ of the particles\az{,} see SI Appendix\az{, SI Text and Fig.~S3}.

Second, we determine $\dot{\gamma}$ of oscillating particles by calculating the angular frequencies $\omega_m$ of the maxima of the Fourier transform of $\psi(t)$.
Using $\omega_m$ and the particle aspect ratios $\alpha$ determined from BEM simulations, we use the relation known from Jeffery dynamics, $\dot{\gamma}=\omega_m(\alpha+\alpha^{-1})$.
The determined shear rates $\dot{\gamma}$ \az{for different flow rates $Q$ and} for different oscillating head-heavy particles of different pitch \az{ is shown in SI Appendix, Fig.~S2}.
The linear relation \az{$\dot{\gamma}=0.9Q~\mathrm{nl^{-1}}$} indeed fits the data well, in agreement with
 the previously described method\az{.}

\subsection*{Boundary Element Method (BEM)}
For the BEM simulations the surface of the spheres and the helical tails are discretized into a triangular mesh with triangle length \az{$\sim 0.075~D$} (\az{SI Appendix, }  Fig.S5).
Similar triangular discretizations have previously been used to model flagellated bacteria, see e.g.\ Refs.\ \cite{Ramia1993,Ishikawa2007c,Shum2010}.
The flow velocity at a given position $\mathbf{x}$ can be obtained by the boundary integral formulation
\begin{equation}
  v_i (\mathbf{x})
  = v_i^\infty (\mathbf{x})
  - \frac{1}{8 \pi \eta} \sum_n^{N_E} G_{ij}(\mathbf{x}, \mathbf{y_n}) q_j (\mathbf{y_n}) dS_n \label{eq:bem}
\end{equation}
where $\mathbf{v}^\infty$ is the background flow, $\eta$ is the viscosity, $N_E$ is the number of mesh triangles, $\mathbf{G}$ is the Oseen tensor, $\mathbf{q}$ is the viscous traction acting at a surface position $\mathbf{y}_n$ and \az{$dS_n$} is the triangle area.
By solving Eq.~(\ref{eq:bem}) together with the constraint of force- and torque-free conditions in a matrix form \cite{Pozrikidis1992,Ishikawa2006}, the translational and rotational velocities of the particle can be obtained.

\section*{acknowledgements}
The authors acknowledge funding from the ERC Consolidator Grant PaDyFlow(Agreement  682367). This work has received the support of Institut Pierre-Gilles de Gennes (\'Equipement d’Excellence, ``Investissements d’avenir'', program ANR-10- EQPX-34).
AZ acknowledges funding from the Austrian Science Fund (FWF) through a Lise-Meitner Fellowship (Grant No M 2458-N36).
DM acknowledges funding from JSPS (Japan Scociety for the Promotion of Science) KAKENHI Grant Number 21H05879 and JST (Japan Science and Technology Agency) PRESTO Grant Number JPMJPR21OA.

\newpage
\setcounter{figure}{0} 
\renewcommand{\thefigure}{S\arabic{figure}}


\noindent {\Large Supplementary Information: \\ Asymmetric bistability of chiral particle orientation in viscous shear flows}\\[4mm]

\subsection{Estimation of local shear rate}
\label{sec:shear}
We estimated the local shear rate $\dot{\gamma}$ experienced by the particles by two independent methods, as discussed in the main text.
The shear rates obtained at different flow rate $Q$ using the second method is shown in Fig.~\ref{fig:S2}.
In the following we present more details regarding the first method.

\az{When we neglect} hydrodynamic particle-wall interactions \az{and the flow curvature,} we \az{can} assume to first approximation that the particles \az{simply} follow the flow velocity $v_x(y,z)$ in the rectangular channel.
For a given channel geometry and applied flow rates $Q$ the flow field $v_x(y,z)$ in the channel can be calculated, with $v_\text{max}$ the velocity in the center of the channel \cite{Bruus2007}.
Since our particles are sufficiently far away from the side walls, the flow profile can be approximated as planar Poiseuille flow, $v_x(z)=(4/h^2)v_{max} (h-z)z$ with $h=96\mu m$.
Measuring the particle velocity and comparing to the flow profile can then be used to extract the particle position $z$ in the channel, and eventually the $z$-dependent shear rates $\dot{\gamma}(z)=dv_x/dz$ \az{of the particles}.
The inset in Fig.~\ref{fig:S1}(a)  shows the distribution of experimental particle velocities normalized with the respective $Q$-dependent $v_{max}$    showing two peaks: 
One peak at $v/v_{max}\approx 0.2$ corresponding to a distance $z \approx R$ away from the wall (indicated by an orange arrow in the inset and main Fig.~\ref{fig:S1}(a)) with $R=D/2$, which correspond to particles rolling at the stable orientations $\psi^\ast=\pm \pi/2$.
    \az{A second peak appears} at $v/v_{max}\approx 0.6$ corresponding to a distance $z \approx l/2=(2R+L)/2 $ (half a particle length) away from the wall (indicated by the green arrow in the inset and main figure), which correspond to particles oscillating with high amplitudes close to the wall. 
In Fig.~\ref{fig:S1}(b) the dependence of the particle velocity on the amplitude  $|\psi_a|$ shows that indeed particles with higher amplitude move faster (i.e. more away from wall) \az{compared to} particles with small amplitude (in particular rolling particles close to the wall). 
    Fig.~\ref{fig:S1}(c) shows theoretical curves demonstrating the relation between
    flow rate $Q$ and local shear rate $\dot{\gamma}$ at different positions $z$. The orange and green arrows point to the two $z$ positions discussed in Fig.~\ref{fig:S1}(a). 
    The larger $z$ (high amplitudes, green arrow) leads to the condition for oscillating particles $\dot{\gamma}\approx 0.9Q/nl$ used in the main text, while rolling particles ($z\sim R$, orange arrow) experience higher shear rates $\dot{\gamma}_+ \approx 1.3Q/nl$.





\subsection{Head-heavy torque}
In our model we take the head-heavy torque of the particle close to the wall into account in a simplified manner.
\az{For large amplitude oscillations the head-heavy torque does not play a significant role for the angular dynamics, as can be seen, for example, in Fig.~5 in the main text, or in comparing large amplitude oscillations in Fig.~4A and Fig.~4B in the main text. It becomes more important when the particles approach their stable positions $\psi^\ast=\pm \pi/2$, again, as can be seen in Fig.~4A,B and Fig 5 in the main text.
Note that for relatively small angle $\psi$ the angle $\theta(t)$ oscillates around zero with even smaller amplitude.
Still, $\theta$ periodically switches between small positive and small negative $\theta$ values.}

\az{When the particle is instantaneously pointing away from the wall, see Fig.~\ref{fig:S3}(a), its head experiences a sedimentation velocity $v_\text{sed}$ due to its heaviness, leading to a head-heavy torque $\Omega_H$ reorienting the particle more parallel to the wall.
However, this is not symmetric in $\theta$: When the particle is instantaneously pointing towards the wall ($\theta<0$), its heavy head is (nearly) touching the wall, and does not experience a sedimentation velocity, and hence the particle does not experience this head-heavy torque, see Fig.~\ref{fig:S3}(b).
Therefore we approximate this effect by the Heaviside step function $\mathcal{H}(\theta)$ as used in Eq.~(2) in the main text.
While this approximation works best for  small amplitude oscillations, where the effect on the dynamics is strongest, it is kept in our model for simplicity for the entire dynamics, in particular also for large amplitudes where head-heaviness is anyhow negligible.}

\subsection{Determination of model parameters with Boundary Element Method in Poiseuille flow}

  \az{The parameters obtained for a particle in shear flow (see also sketch in Fig.~\ref{Fig:BEM-Sketch}(a)) as described in the main text using the Boundary Element Method (BEM) leads to well-defined particle chiral strength $\nu$, effective aspect ratio $\alpha$, and bottom-heavy strength $\Omega_H$, independent of the channel geometry. In the following we also measure the values of $\nu$ and $\alpha$ in Poiseuille flow, both (i) with and (ii) without the presence of bounding walls (see also sketches in Fig.~\ref{Fig:BEM-Sketch}(c,d)), to test the applicability of using the simple shear flow approximation to describe our experimental setup.
  To include the hydrodynamic interactions between the particle and the wall we use the conventional Blake solution for the Green's function of flow singularities near a wall, and we add up the contributions from both walls, similar as in previous work, see e.g.\ Ref~\cite{Matsunaga2017}.}
  
  \az{Figure \ref{Fig:BEM-PF}(a) shows the angular velocity $\Omega_y$ which determines the particle aspect ratio $\alpha$, as discussed above, for the case of Poiseuille flow with walls. The dashed lines are the reference values obtained in simple shear flow without walls. The values are measured in units of the shear rates $\dot{\gamma}$ at the particle positions. We can see that the rotation rates reduce when getting closer and closer to the bottom wall, similar as observed for ellipsoids in flow near a wall \cite{Pozrikidis2005,Kaya2009}. The effect is relatively weak for high-amplitude oscillations where the particle is expected to be relatively far away from the wall (see also discussion in Sec.~\ref{sec:shear}), about half its length or $l_z \sim 6R$.
  The corresponding angular velocities $\Omega_z$ are shown in Fig.~\ref{Fig:BEM-PF}(b), which determine the actual chiral reorientation strength in the microchannel Poiseuille flow. It can be seen that the effect of the quadratic flow and the walls is almost negligible unless the particle gets very close to the wall, where the chiral strength slightly increases.}

  \az{In Figures \ref{Fig:BEM-PF}(c,d) we compare $\Omega_y$ and $\Omega_z$ in Poiseuille flow with walls (light blue curves) to the results in Poiseuille flow without walls (dark blue curves) and to the results in simple shear without walls (black dashed lines). We can clearly see that deviations from the simple-shear case only results from the wall interactions, while the effect from the quadratic flow profile can be neglected. This can be understood by the fact that the shear rate varies linearly, and the relevant shear rate the particle experiences is to a good approximation the mean shear rate, at the center of the particle. All in all we see that the effects of the wall and the quadratic flow profile is small for the particle dynamics, and  particularly small for the chiral reorientation, quantified by the chiral strength $\nu$.}

\subsection{Supplementary Figures amplitude decay}
In Figs.~\ref{fig:S5}-\ref{fig:S7} we show the data for the amplitudes of all trajectories for all experimentally studied particle shapes and shear rates.
Individual amplitude maxima $|\psi_a|$ are shown as single points. Each subfigure typically consists of multiple trajectories.
In Fig.~\ref{fig:S5} we show data for the head-heavy particles, in Fig.~\ref{fig:S6} data for the shell particles.
In Fig.~\ref{fig:S7} we show the amplitudes for the double-helix particles. The results are in accordance with the model results shown in Fig.~4 in the main text, i.e.\ the classification into (almost) not decaying amplitudes at very high initial amplitudes, and the decay due to head-heaviness for smaller initial amplitude.

\subsection{Movie Description}
In movies S1-S6 the dynamics described in Fig.~2 in the main text is shown. In all movies the flow goes from right to left, and the images obtained from the moving camera are centered around the center of the particle.

\begin{itemize}
\item{Movie S1: Stabilization of left-handed particle of pitch $p=5\mu m$ at $\psi^\ast=\pi/2$ obtained at flow rate $Q=30nl/s$.}
\item{Movie S2: Flipping and stabilization of left-handed particle of pitch $p=5\mu m$ at $\psi^\ast=\pi/2$ obtained at flow rate $Q=30nl/s$.}
\item{Movie S3: Stabilization of left-handed particle of pitch $p=5\mu m$ at $\psi^\ast=-\pi/2$ obtained at flow rate $Q=10nl/s$.}
\item{Movie S4: Stabilization of right-handed particle of pitch $p=5\mu m$ at $\psi^\ast=-\pi/2$ obtained at flow rate $Q=30nl/s$.}
\item{Movie S5: Flipping and stabilization of right-handed particle of pitch $p=5\mu m$ at $\psi^\ast=-\pi/2$ obtained at flow rate $Q=30nl/s$.}
\item{Movie S6: Stabilization of right-handed particle of pitch $p=5\mu m$ at $\psi^\ast=\pi/2$ obtained at flow rate $Q=10nl/s$.}
\end{itemize}


%



\begin{figure*} [tb]
\centering
    \includegraphics[width=7cm]{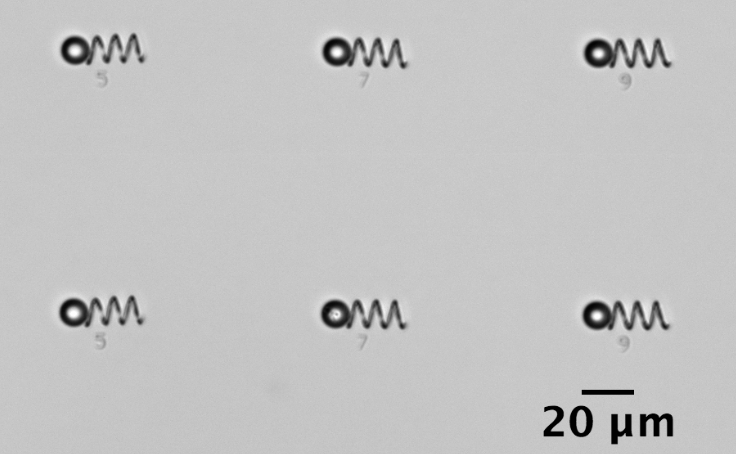}
    \caption{\az{Example of array of identical particles printed on the substrate.}
    }
    \label{fig:S0}
\end{figure*}

\begin{figure*} [tb]
\centering
    \includegraphics[width=.35\textwidth]{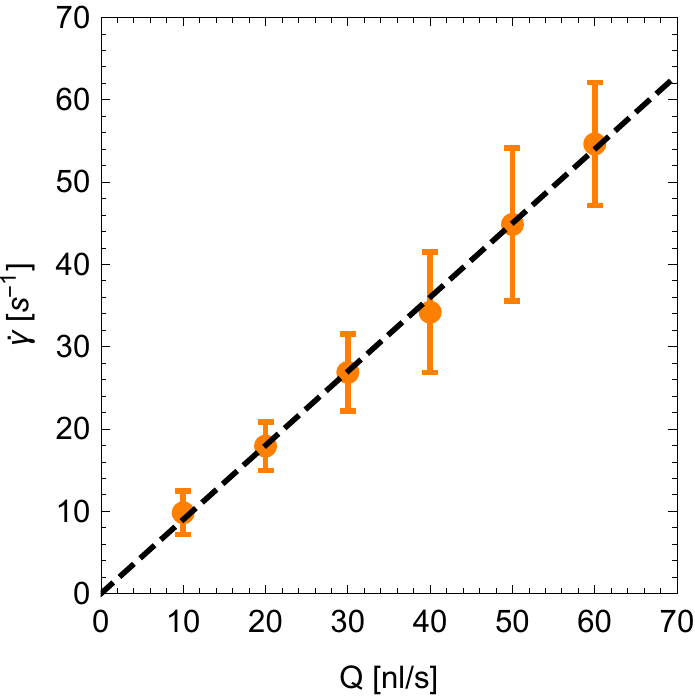}
    \caption{Shear rates $\dot{\gamma}$ of particles estimated from the peaks of the power spectra of $\psi(t)$, averaged over all oscillating head-heavy particle trajectories at a given flow rate $Q$. 
    The black dashed line shows the relation $\dot{\gamma}=0.9Q/nl$.
    }
    \label{fig:S2}
\end{figure*}

\begin{figure*} [tb]
\centering
    \includegraphics[width=\textwidth]{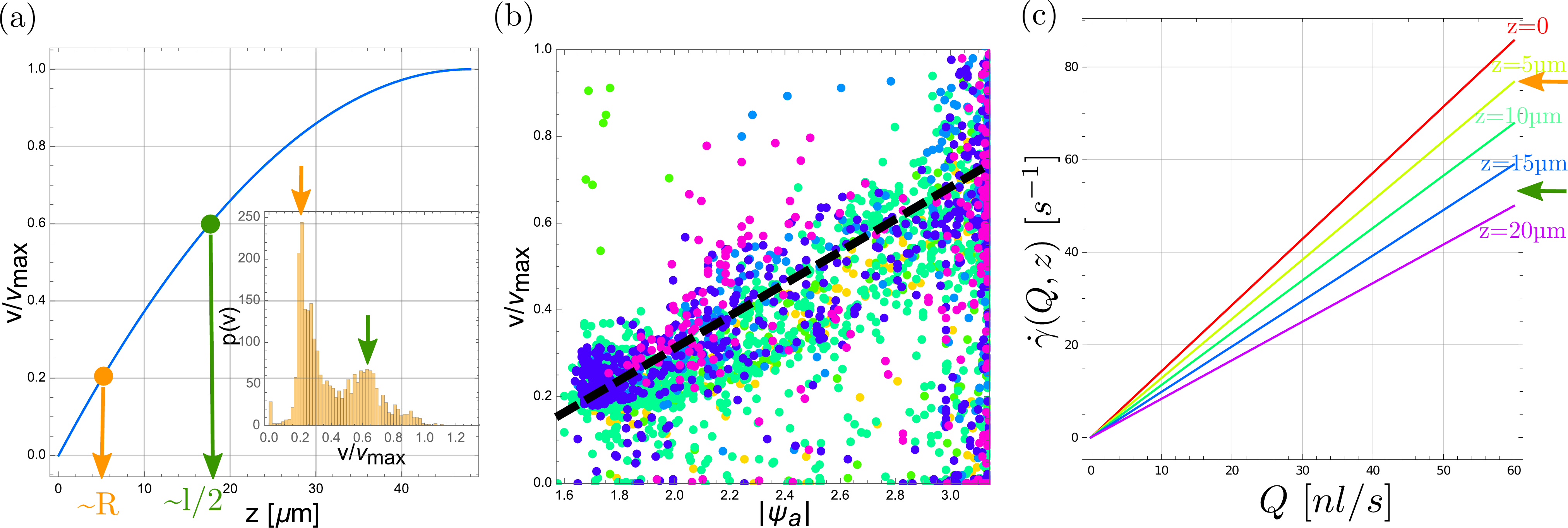}
    \caption{Relation between particle velocity $v_x(z)$, particle height $z$, and shear rate $\dot{\gamma}$ (for details see Supplementary text).
    (a) Relation between particle position $z$ and velocity $v_x$.
    (b) Dependence of \az{instantaneous} particle velocity on the \az{instantaneous} amplitude height $|\psi_a|$. Shown is \az{the experimental} data for all particle shapes and all shear rates.
    (c) Theoretical curve demonstrating the relation between
    flow rate $Q$ and local shear rate $\dot{\gamma}$ at different positions $z$. The orange and green arrows point to the two $z$ positions discussed in (a). 
    The larger $z$ (high amplitudes, green arrow) corresponds to oscillating particles 
    with the relation $\dot{\gamma}=0.9Q/nl$.
    }
    \label{fig:S1}
\end{figure*}

\begin{figure*} [tb]
\centering
    \includegraphics[width=.6\textwidth]{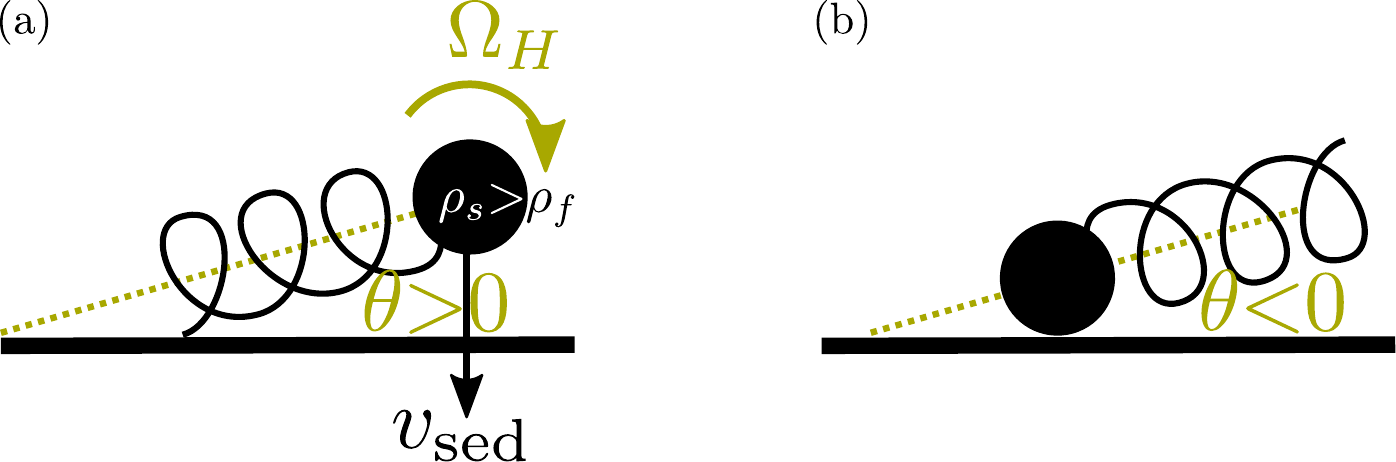}
    \caption{\az{Sketch of the head-heavy torque acting on the particle. (a) The head is not touching the wall ($\theta > 0$) where the sphere-sedimentation velocity $v_{sed}$ leads to a torque $\Omega_H$ reorienting the head-heavy particle pointing head-down. (b) When the head is touching the wall ($\theta<0$) head sedimentation and hence the corresponding torque is suppressed.}}
    \label{fig:S3}
\end{figure*}

\begin{figure*} [tb]
\centering
    \includegraphics[width=\textwidth]{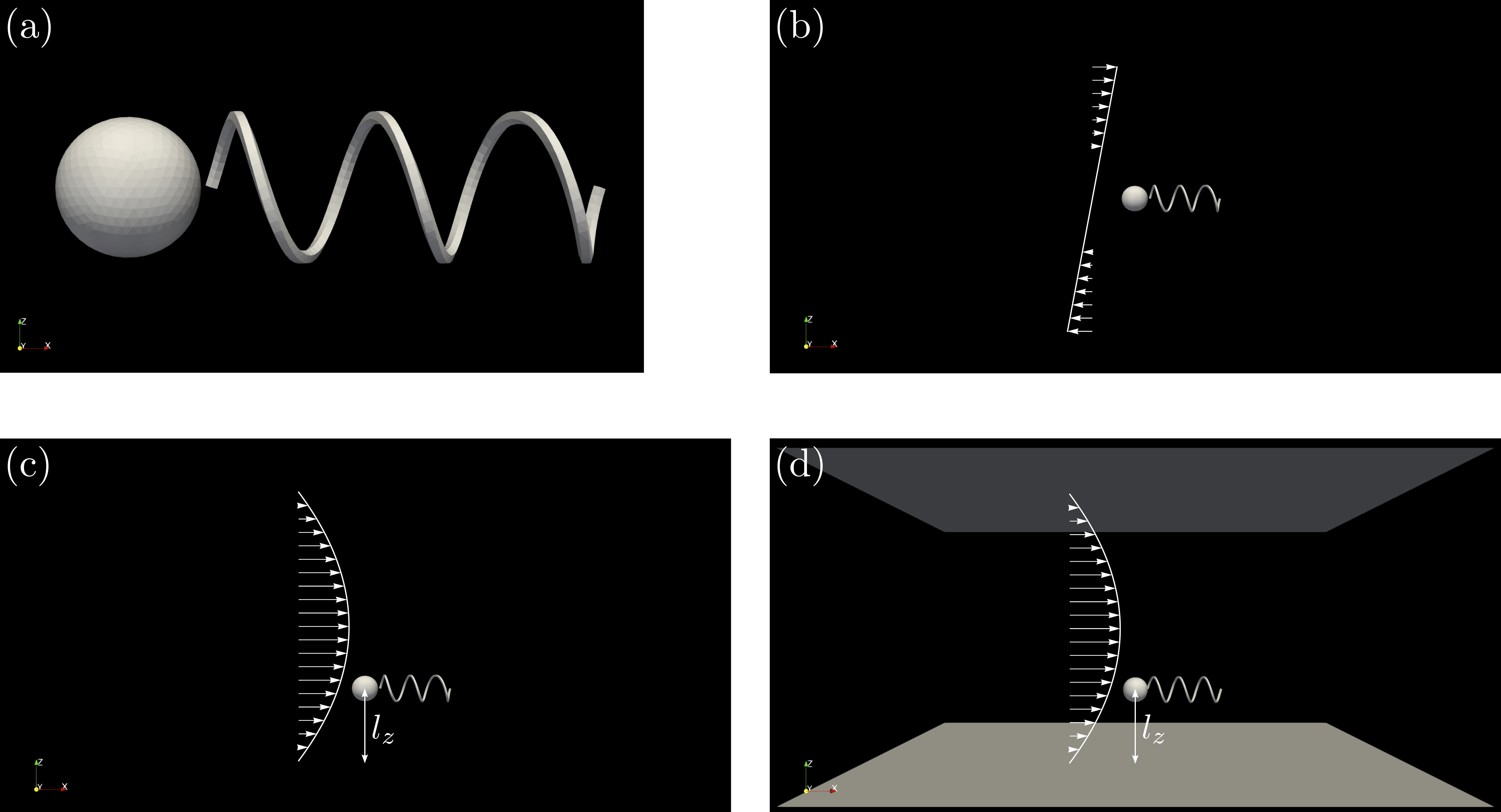}
    \caption{\az{(a) Sketch of a triangulated particle used in the BEM simulations, shown here for pitch $p=10\mu m$ and $\chi=+1$. (b) Sketch of particle aligned in linear shear flow which was used to calculate the parameters $\nu$ and $\alpha$ used in the main text. (c,d) particle aligned in planar Poiseuille flow (c) without and (d) in the presence of walls. The distance of the particle from the bottom wall is denoted by $l_z$.}}
    \label{Fig:BEM-Sketch}
\end{figure*}


\begin{figure*} [tb]
\centering
    \includegraphics[width=.8\textwidth]{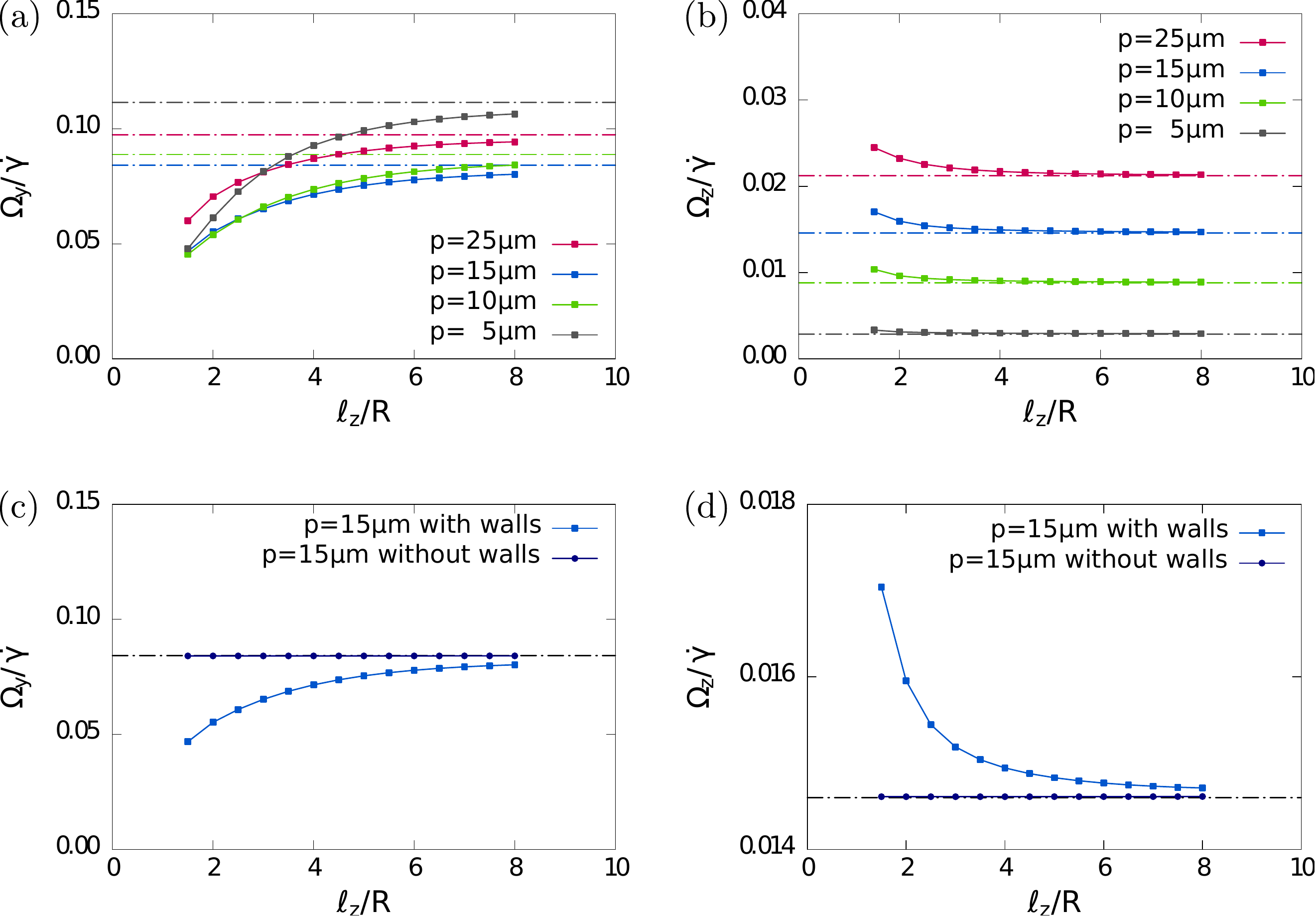}
    \caption{\az{Results of instantaneous angular velocities $\Omega_y$ and $\Omega_z$ normalized by the local shear rate $\dot{\gamma}$ from BEM simulations for a particle aligned in Poiseuille flow (see also sketch in Fig.~\ref{Fig:BEM-Sketch}(d)), compared to the corresponding results in simple shear without walls (dashed lines). (a,b) Results for different particle shapes at different distances $l_z$ away from the bottom wall, normalized by the head radius $R$. (c,d) Results for particle with pitch $p=15\mu m$ in Poiseuille flow in the presence of walls compared to Poiseuille flow without walls (see also sketch in Fig.~\ref{Fig:BEM-Sketch}(c)). Note the different scale of the $y$-axis in (d).}
    }
    \label{Fig:BEM-PF}
\end{figure*}

\begin{figure*} [tbh]
\flushright
    \includegraphics[width=\textwidth]{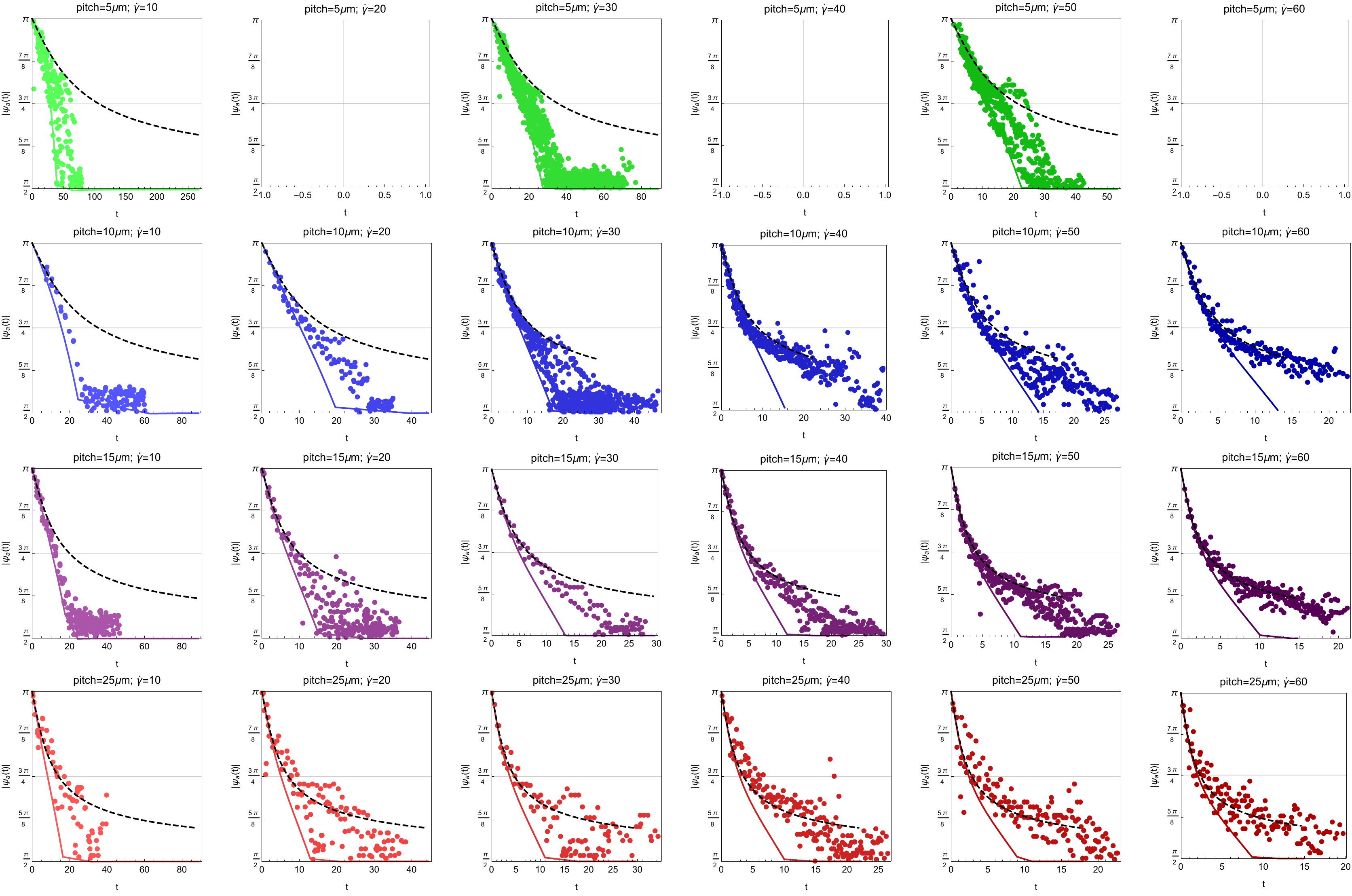}
    \caption{Amplitude decay for head-heavy particles depending on particle pitch and flow rate $Q$. Note that for some combinations of $Q$ and pitch no  experimental data exists.
    The solid lines are the solutions from the theoretical model (Eqs.~(1) and (2) with full bottom-heaviness from main text). The black dashed lines are the solutions without bottom-heaviness.}
    \label{fig:S5}
\end{figure*}

\begin{figure*} [bth]
\flushright
    \includegraphics[width=.833\textwidth]{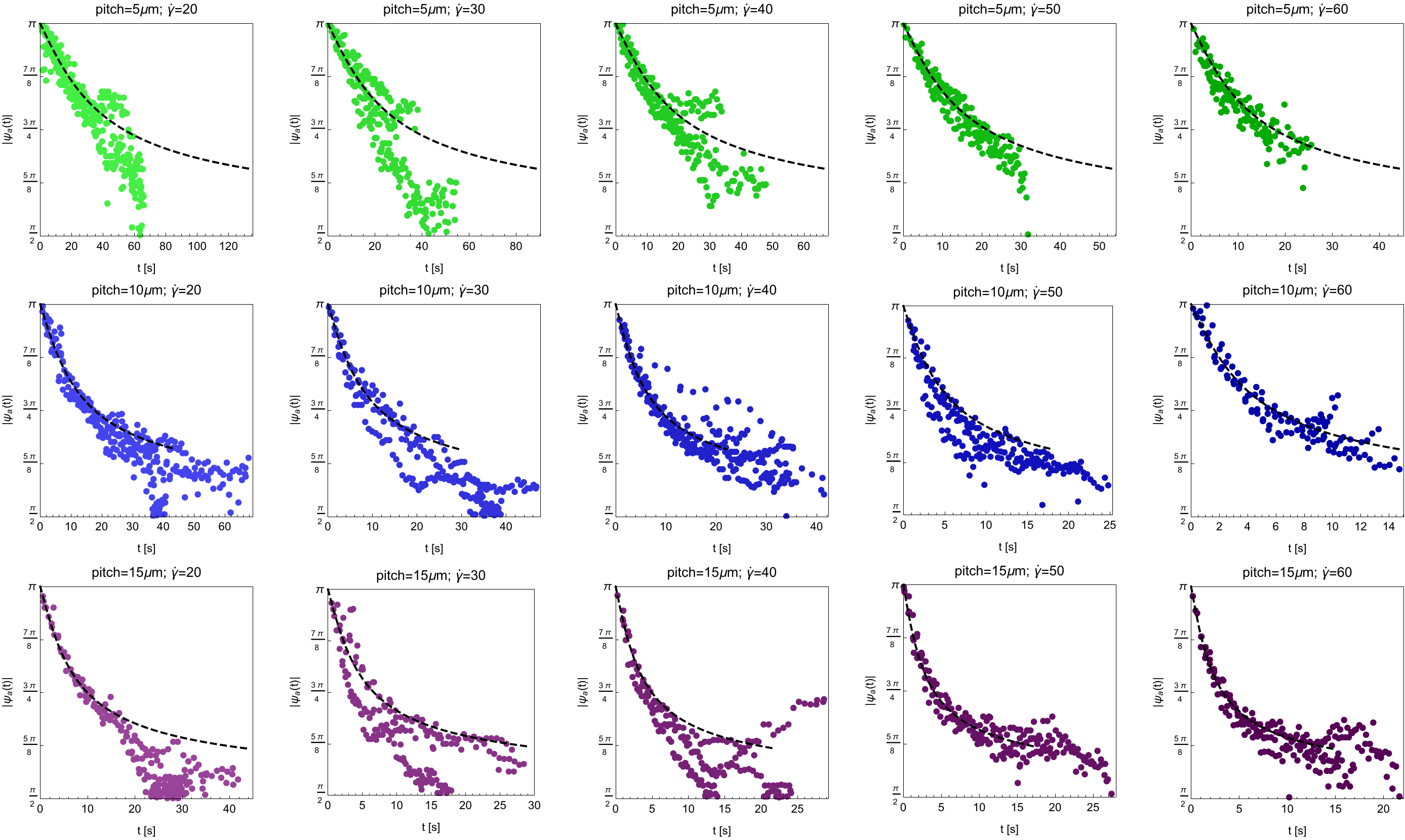}
    \caption{Amplitude decay for shell particles depending on particle pitch and flow rate $Q$. Note that for some combinations of $Q$ and pitch no  experimental data exists. The black dashed lines are the solutions from the theoretical model (Eqs.~(1) and (2) from main text) without bottom-heaviness.}
    \label{fig:S6}
\end{figure*}

\begin{figure*} [htb]
\centering
    \includegraphics[width=.4\textwidth]{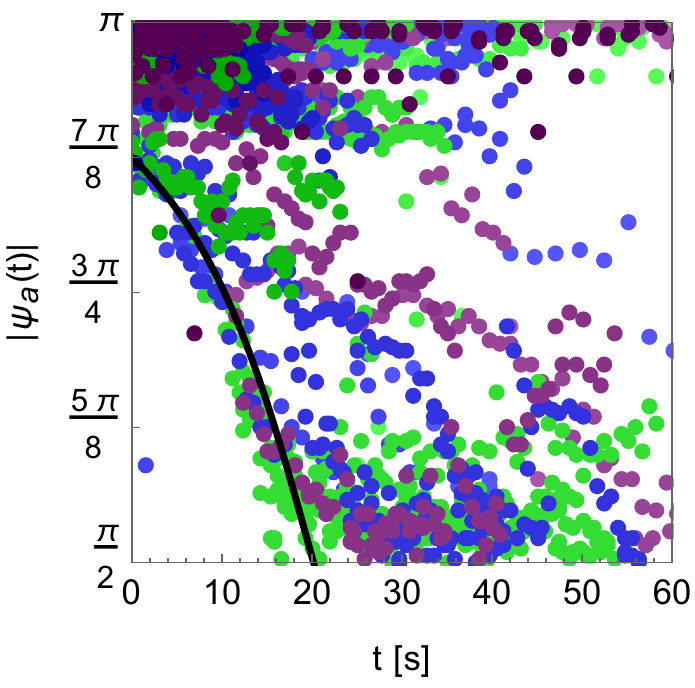}
    \caption{Amplitude decay of double helix particles.
    Results are in accordance with theoretical solution for $\nu=0$ shown in
    Fig.~4C in the main text:
    Non-chiral particles (i) either perform oscillations at high amplitudes
    without significant decay, (ii) or for smaller initial amplitude decay towards the stable positions $\psi^\ast=\pm \pi/2$ due to head-heaviness.
    The black line shows the theoretical curve for initial $\psi_0=7\pi/8$
    which is the same for all shear rates. 
    }
    \label{fig:S7}
\end{figure*}


\end{document}